\documentclass[11pt,a4paper]{article}
\usepackage{latexsym,amsfonts,graphicx}

\newtheorem{thm}{Theorem}
\newtheorem{lem}{Lemma}
\newtheorem{cor}{Corollary}
\newtheorem{prop}{Proposition}

\newcommand{\cL}{{\mathcal L}}
\newcommand{\1}{\mathbb{I}}
\newcommand{\U}{{\sf U}}

\newcommand{\Def}{{\mathcal N}}

\newcommand{\R}{\mathbb{R}}
\newcommand{\C}{\mathbb{C}}

\newcommand{\mb}{\beta}
\renewcommand{\ll}{{\lambda_0}}

\newcommand{\diag}{\mbox{diag}}
\newcommand{\Kl}{{\mathcal Q}_0}
\newcommand{\Rl}{{\mathcal R}_0}

\renewcommand{\P}{{\mathcal P}}

\title{Manipulating the  electron current through a splitting}
\author{M. Harmer, A. Mikhailova and B. S. Pavlov}

\date{}

\begin{document}

\maketitle

\begin{abstract}
The description of electron current through a splitting is
a mathematical problem of electron transport in quantum networks
\cite{Exn,Ada}. For quantum networks constructed on the interface of 
narrow-gap  semiconductors \cite{YBR,APY} the relevant scattering
problem for the multi-dimensional Sch\"{o}dinger equation  may  be 
substituted by the corresponding problem on a one-dimensional linear
graph with proper selfadjoint  boundary conditions at the nodes
\cite{Gera:Pav,Gera,Nov2,Nov3,Kost:Sch,Mel,Car2,Soa,Mel:Pav,Kur1,Exn:Seb,Exn,
Ada}. However, realistic  boundary conditions for splittings have not
yet been derived. \\ 
Here we consider some compact domain attached to a few semi-infinite
lines as a model for a quantum network. An asymptotic formula for the
scattering matrix for this object is derived in terms of the properties
of the compact domain. This allows us to propose designs for devices for 
manipulating quantum current through a splitting \cite{BMPY,Har2,MPPRY,
GPP,Mik:Pav}.
\end{abstract}

\section*{Introduction: current manipulation in the resonance case} 
In this paper we discuss the scattering problem on a compact domain with a
few semi-infinite wires attached. This is motivated by the design of
quantum electronic devices for triadic logic. In the papers
\cite{BMPY,Har2} a special design of the 
one-dimensional graph which permits manipulation of the current through
an elementary  ring-like splitting is suggested. This permits, {\it in 
principle},  manipulation of  quantum current  in the resonance  case to
form a quantum  switch. Another device for manipulating quantum current
through splittings is  discussed in \cite{MPPRY, GPP}. In \cite{Mik:Pav}
the special design of the splitting formed as a  circular domain with
four one-dimensional wires attached is used to produce a triadic relay. \\
In order to illustrate the basic principle of operation consider the
self-adjoint Schr\"{o}dinger operator
$$
\left\{ 
\begin{array}{c}
\cL \equiv -\Delta + q(x) \\
\left.\frac{\partial \Psi}{\partial n }\right|_{\partial\Omega} = 0 .
\end{array} \right.
$$
on some compact domain $\Omega$. In this paper we will only consider the
case $\Omega\subset\R^3$ \cite{MPPRY} (for other cases see also
\cite{BMPY, Har2, Mik:Pav}). Roughly speaking the solution of the
Cauchy problem 
\begin{equation}
\left\{ 
\begin{array}{rcl}
\frac{\partial \Psi}{i\partial t } & = & \cL \Psi \\
\Psi (x,0) & = & \Psi_0 (x)
\end{array} \right.
\end{equation}
is given in terms of eigenfunctions $\varphi_n$
$$
\Psi (x,t)= \sum_{n} \alpha_{n}e^{i\lambda_n t} \varphi_n (x) .
$$
Picking a specific mode $\varphi_0$ with energy $\lambda_0$ we
suppose that $\varphi_0$ disappears on some subset $l_0\subset\Omega$.
\begin{figure}[ht]
\begin{center}
\includegraphics[height=2.5in, width=2.8in]{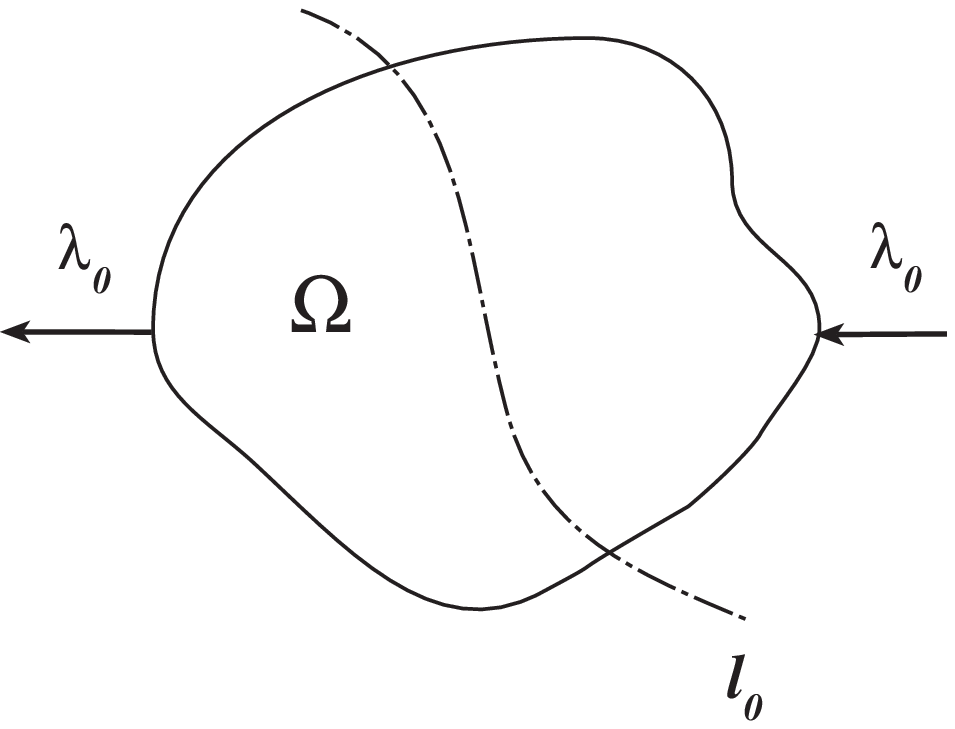}
\\
figure 1. Resonance switch 
\end{center}
\end{figure}
Connecting `thin channels' at various points on the boundary
of $\Omega$ and introducing an excitation of energy $\lambda_0$ along the
channels we can hope to create a switching effect. Essentialy this is achieved 
by varying $q(x)$ so that $l_0\cap\partial\Omega$ coincides with the connection 
point of a `thin channel'. \\
Implicit in our construction is the assumption that the energy of the electrons 
in the device is equal to some resonance eigenvalue of the Schr\"{o}dinger 
operator on $\Omega$. We refer to this as the {\em resonance 
case}\footnote{This has interesting implications when we consider the effect of 
decreasing the length scale---or equivalently scaling up the energy---viz. the 
effect of non-zero temperature becomes negligable for sufficiently small length 
scales, see \cite{Har}.}. \\ Another assumption which we have made above is 
that $\lambda_0$ is a {\em simple} eigenvalue of $\cL$. We will show that the 
case of multiple eigenvalues is a simple generalisation of the case for simple 
eigenvalues, see \cite{BMPY, Har2}. \\ 
In the first section we give a brief description of the connection of the thin 
channels (here they are modelled by one-dimensional semi-lines) to the compact 
domain, for more details see \cite{MPPRY}. In the second section we derive an 
asymptotic formula for the scattering matrix in terms of the eigenfunctions on 
the compact domain. In the last section we briefly discuss some simple models 
of a quantum switch constructed on the basis of this asymptotic formula. \\

\section{Connection of compact domain to thin channels} 

As we mentioned above the thin channels are modelled by one-dimensional 
semi-lines. This is justified by an appropriate choice of materials 
(narrow-band semiconductors) and energies, see \cite{YBR, APY, Mik:Pav}. We 
assume that these channels are attached at the points $\{ a_1,\,a_2,\,...,a_N\} 
\subset \partial \Omega$ (perturbation of the operator $\cL$ at inner points 
$\{a_{N+1},a_{N+2},\ldots$ $,a_{N+M}\}$ may be considered using the same 
techniques as for $\{ a_1,\,a_2,\,...,a_N\}$ \cite{APY, BMPY, Mik:Pav} although 
we do not consider this here). \\ 
We refer to $\cL$, defined above, as the {\em unperturbed} Schr\"odinger 
operator. $\cL$ is restricted to the symmetric operator $\cL_0$ defined on the 
class $D_0$ of smooth functions with Neumann boundary conditions which vanish
near the points $a_1, a_2,\ldots ,a_N$. The deficiency subspaces, $\Def_{\pm 
i}$, of the restricted symmetric operator $\cL_0$, $$
\left[ \cL^{\star}_0 \pm i \right] e^s_{\pm i} = 0  $$
for complex values of the spectral parameter $\lambda$ coincide with Greens 
functions $G_{\lambda}(x,a_s)$ of $\cL$ which  are  elements  of $L_2 (\Omega)$ 
but  do  not  belong  to the  Sobolev class $W_2^1 (\Omega)$. In the case when 
$\Omega$ is a compact one-dimensional manifold (a compact graph) these Greens 
functions are continuous and can be written in terms of a convergent spectral 
series \cite{BMPY}. However, when $\Omega\subset\R^2,
\R^3$ the deficiency elements will have singularities and we must use 
an iterated Hilbert identity to regularise the values of the Greens function at
the poles. \\ It is well known, for $\Omega\subset\R^3$, that the 
Greens function admitts the representation inside $\Omega$
\begin{equation}\label{formula} G^0_{\lambda} (x,y) = 
\frac{e^{i\sqrt{\lambda}|x-y|}}{4\pi|x-y|} + g(x,y,\lambda)  
\end{equation} where $g(x,y,\lambda)$ is continuous. The  potential-theory  
approach gives the asymptotics  of the Green function  near  the  boundary  
point  $a_s\in\partial \Omega$ 
\begin{equation}\label{3} G^0_{\lambda}(x,a_s) \sim \frac{1}{2\pi|x-a_s|} + 
L_s(x) +  B_s(x,\lambda) , \end{equation}
where $L_s$ is a logarithmic term depending only on $\partial\Omega$ and $B_s$ 
is a bounded term containing spectral information \cite{MFad}. \\ In order to 
choose regularised boundary values we use the following lemma \cite{MPPRY} 
(here we assume $\cL > -1$ is semi-bounded from below): \begin{lem}
For any regular point $\lambda $ from the  complement of the  spectrum 
$\sigma(\cL)$ of $\cL$ and  any  $a \in \left\{a_s\right\}_{s=1}^{N+M}$ the 
following representation is true:
$$ G_{\lambda}(x,a) = G_{-1}(x,a) + (\lambda + 1) G_{-1}*G_{\lambda} (x,a),
$$ where the second addend is a continuous function of $x$ and the 
spectral series of it in terms of eigenfunctions $\varphi_l$ of the 
nonperturbed operator $\cL$ $$
(\lambda + 1) G_{-1}*G_{\lambda} (x,a) = (\lambda + 1) \sum_{l} \frac{\varphi_l 
(x) \varphi_l (a)}{(\lambda_l + 1) (\lambda_l - \lambda)}
$$ is absolutely and uniformly convergent in $\Omega$. 
\end{lem} The proof of this lemma is based on the classical Mercer theorem 
along with the Hilbert identity \cite{MPPRY}. \\
It is well known that the domain of $\cL^{\star}_{0}$ can be written as 
the direct sum
\begin{equation}\label{vNi} D^{\star}_0 = D_0 +  \Def_i + \Def_{-i} 
\end{equation} so for any $u\in D^{\star}_0$
$$ u  = u_0 + \sum_s A^+_s G_i (x,a_s) + \sum_s A^-_s G_{-i} (x,a_s) .
$$ We define $u\in D^{\star}_0$ in terms of the coordinates
\begin{eqnarray*} A_s & \equiv & A_s^+ + A_s^- , \\
B_s & \equiv & \lim_{x\to a_s} \left[ u(x) - \sum_t A_t \Re G_i (x,a_t) \right] 
, \end{eqnarray*}
the {\em singular} and {\em regular} amplitudes respectively since it is clear 
from the above lemma that $A_s$ is the coefficient of the singular part and 
$B_s$ the coefficient of the regular part of $u\in D^{\star}_0$. The boundary 
form of $\cL^{\star}_0$ may be written in terms of
$A_s$ $B_s$ as a hermitian symplectic form
\begin{equation}\label{A_s,B_s} \langle \cL^{\star}_{0} u,v \rangle - \langle 
u,\cL^{\star}_{0} v \rangle = \sum B^u_s \bar{A}^v_s - A^u_s \bar{B}^v_s .
\end{equation}

\subsection{Self-adjoint extensions}

We recall that to each boundary point $a_s$, $s=1,\ldots
,N$, there is attached a semi-infinite ray. On the $s$-th ray we define the
symmetric operator 
$$
l_{s,0} = -\frac{d^2}{dx_s} + q_s(x_s),
$$
on functions which vanish at $x_s =0$ (which is identified with
$a_s\in\Omega$). \\ 
Let us consider the symmetric operator $\cL_{0} \oplus
l_{1,0} \oplus l_{2,0} \oplus...\oplus l_{N,0}$. The connection between
the compact domain and the rays is given by (a particular) self-adjoint
extension of this operator.  The boundary form of the adjoint
$\cL^{\star}_{0} \oplus l_{1,0}^{\star}
\oplus l_{2,0}^{\star} \oplus...\oplus l_{N,0}^{\star}$ is easily seen to be
\begin{equation}\label{bform}
\sum^{N}_{s} \left( B^u_s \overline {A^v_s} - A^u_s\overline {B^v_s}
\right) + \sum^{N}_s \left( u^{\prime}_s(0) \overline{v_s}(0) -
u_s(0)\overline{v^{\prime}_s}(0) \right).
\end{equation}
It is well known that the self-adjoint extensions of $\cL_{0} \oplus l_{1,0}
\oplus l_{2,0} \oplus...\oplus l_{N,0}$ correspond to Lagrange
planes in the Hermitian symplectic space of boundary values equipped
with the above boundary form \cite{Pav}. In general, if $A$, $B$ are
(vectors of) boundary values for some symmetric operator then any
self-adjoint extension can be described by
$$
\frac{i}{2}(U - \1 ) A + \frac{1}{2}(U + \1 ) B = 0 
$$
for some unitary matrix $U$ \cite{Har, Har3}. \\
We choose the particular family of self-adjoint extensions which
correspond to the following boundary conditions at the
$N$ points of contact of the rays
\begin{equation}\label{6b}
\left( \begin{array}{c}
A_s \\
u_s (0) 
\end{array} \right) = \left( \begin{array}{cc}
0     & \beta \\
\beta & 0 
\end{array} \right)
\left( \begin{array}{c}
B_s \\
u'_s (0) 
\end{array} \right) ,
\end{equation}
for $s=1,\ldots ,N$ and $\beta > 0$. The resulting self-adjoint extension
we denote by $\cL_{\beta}$. The parameter $\beta$ is a measure of
the strength of the connection between the rays and the compact
domain---in the limit
$\beta\rightarrow 0$ the resolvent of $\cL_{\beta}$ converges uniformly to
the resolvent of $\cL$ on each compact subset of the resolvent set of 
$\cL$ \cite{MPPRY}. 

\section{Asymptotics of the scattering matrix} 

For the remainder we assume that the potential on the rays $q_s  (x_s
)\equiv 0$ is zero. \\
We use the ansatz 
\begin{equation}
\label{Jost}
 u_s = f_s(x_s,-k) \delta_{s1}+ f_s(x_s,k) S_{s1},
\end{equation}
for the scattered wave generated by the incoming wave from the ray
attached to the point $a_1$. Here $f_s(x_s,\pm k)$ are the Jost 
solutions \cite{Ree:Sim}, in this case ($q_s  (x_s
)\equiv 0$) just the exponentials
$$
f_s(x_s,\pm k) = e^{\pm ikx_s} ,
$$ 
and $\lambda=k^2$ is the spectral parameter. \\
>From the boundary conditions (\ref{6b}) we get $N$ equations
\begin{eqnarray}\label{bcl}
A_s & = & \beta f^{\prime}_s(0,-ik) \delta_{s1} + \beta
f^{\prime}_s (0,ik) S_{s1} \nonumber \\
\beta B_s & = & f_s(0,-ik) \delta_{s1} + f_s(0,ik) S_{s1} .
\end{eqnarray}
Inside $\Omega$ the eigenfunction $u(x,k)$ may be written as a sum of Greens
functions at the spectral parameter $\lambda=k^2$
$$
u(x,k) = \sum^{N}_{s} C_s G_{\lambda} (x,a_s ) .
$$
Using the Cayley transform between the spectral points $i$ and $\lambda$ one 
gets a relationship between these Greens functions and the deficiency
elements (as defined above) so that \cite{MPPRY}
$$
\lim_{x \rightarrow a_s}\left[G(x,a_s,\lambda) - \Re
G(x,a_s,i)\right] = \left( \frac{\1 + \lambda \cL }
{\cL - \lambda\1} G_i (a_s), G_i (a_s)\right) \equiv g^s(\lambda) .
$$
Consequenttly we can show that $u$ has the following asymptotics as
$x\rightarrow a_s$
\begin{equation}\label{5}
u \sim C_s \Re G_{i}(x,a_s) + C_s g^s(\lambda) + \sum_{t \neq s} C_t
G_{\lambda}(a_s,a_t) + o(1) .
\end{equation}
It follows that for the scattering wave the symplectic variables are
related by
\begin{eqnarray*}
A_s & = & C_s \\
B_s & = & g^s (\lambda) C_s + \sum_{t \neq s} C_t
G_{\lambda}(a_s,a_t) ,
\end{eqnarray*}
that is $B=QA$ where
\begin{equation}\label{Qmatrix}
Q (\lambda) = \left( \begin{array}{cccc}
g^1(\lambda) & G_{\lambda}(a_1,a_2) & \cdots & G_{\lambda}(a_1,a_{N}) \\
G_{\lambda}(a_2,a_1) & g^2(\lambda) & \cdots & G_{\lambda}(a_2,a_{N}) \\
\vdots               &              & \ddots & \vdots \\
G_{\lambda}(a_{N},a_1) &  \cdots  & \cdots & g^{N}(\lambda)
\end{array}
\right) . 
\end{equation}
Putting this into (\ref{bcl}) we can solve for the scattering matrix to get
\begin{equation}\label{7}
S = - \frac{\1 + ik\beta^2 Q }{\1 - ik\beta^2 Q} .
\end{equation}
Let us choose an eigenvalue $\lambda_0$ of the unperturbed
operator $\cL$ on $\Omega$. We suppose that $\lambda_0$ has a
$p$-dimensional eigenspace, which we denote $\Rl$, with orthonormal basis
$\{\varphi_{0,i}\}^p$. The following important technical statement
close to Lemma 1 above is true \cite{MPPRY}:
\begin{thm}
The elements of the $Q$-matrix have the following asymptotics at
the spectral point $\lambda_0$:
$$
Q_{s t}(\lambda) \sim \sum^p_{i=1}\frac{\varphi_{0,i} (a_s)\varphi_{0,i}
(a_t)}{\lambda_0 - \lambda} + \Kl (a_s,a_t,\lambda) ,
$$
where $\Kl (a_s,a_t,\lambda)$  is  a   continuous  function  at  the 
point $\lambda = \lambda_0$.
\end{thm}
We will use this result to prove an asymptotic formula for the scattering 
matrix in the limit of weak connection between the compact domain and the rays. 
\\ Consider the mapping $\P :L_2(\Omega) \rightarrow \C^N$ which
gives the vector of values of a function in $L_2(\Omega)$ at the nodes of each 
of the $N$ rays. To distinguish between functions and elements of $\C^N$ we use 
the notation $$
\P(\psi)=|\psi\rangle\in\C^N , $$
and we denote $$
R_0 \equiv \P (\Rl) . $$
\begin{prop} It is possible to choose an orthonormal basis $\{\phi_{0,i}\}^p$ 
for $\Rl$ which forms an {\em orthogonal}, but not necessarily normalised,
basis for $R_0$ under $\P$. \end{prop}
{\it Proof:} Given some orthonormal basis $\{\varphi_{0,i}\}^p$ for $\Rl$ we 
see that  $$
\phi_{0,i}=\sum^p_{j=1} U_{ij}\varphi_{0,j} $$
is also an orthonormal basis where $U\in\U (p)$. \\ The inner product of the 
image under $P$ $$
\langle\phi_{0,i}|\phi_{0,j}\rangle=\sum^p_{r,s=1} 
\bar{U}_{ir} \langle\varphi_{0,r}|\varphi_{0,s}\rangle U_{js}
$$ shows that finding an {\em orthogonal} basis for $\Rl$ amounts to finding
the unitary matrix $U$ which 
diagonalises $A_{rs}=\langle\varphi_{0,r}|\varphi_{0,s}\rangle$. 
\hspace*{\fill} $\Box$ \\

This allows us to write $Q$ in `diagonal' form
\begin{eqnarray}\label{Gf}
Q & = & \frac{1}{\ll-\lambda} \left[ |\phi_{0,1}\rangle
\langle\phi_{0,1}|+\cdots+ |\phi_{0,m}\rangle
\langle\phi_{0,m}| \right] \nonumber \\
& & \mbox{} + \Kl (\lambda) \nonumber \\ 
& = & \frac{D_l}{\ll-\lambda} + \Kl (\lambda)
\end{eqnarray}
where $m\le p$ is the dimension of $R_0$.
\begin{thm}\label{fkff}
If $\ll$ is an eigenvalue of $\cL$ then for vanishing
coupling $\mb\sim 0$ the scattering matrix of $\cL_{\beta}$ has the form
\begin{eqnarray}\label{asym}
S(\ll) & = & -\1 + 2P_0 - 2\sum_{s=1}(ik_0\mb^2 P_0^{\perp}
\Kl P_0^{\perp})^s \nonumber \\
& = & -\1 + 2 P_0 + O(\mb^2)
\end{eqnarray}
where $P_0$ is the orthogonal projection onto $R_0$.
\end{thm}
{\it Proof:} 
Using equation (\ref{Gf}), 
$$
S(\lambda) = - \left[\1+\frac{ik\mb^2
D_0}{\ll-\lambda}+ik\mb^2\Kl\right] \left[\1-\frac{ik\mb^2
D_0}{\ll-\lambda}-ik\mb^2\Kl\right]^{-1} .
$$
Since $D_0=D_0^{\star}$, the matrix
$E_0 =\1-\frac{ik\mb^2 D_0}{\ll-\lambda}$ is invertable. 
Consequently the denominator can be written
\begin{eqnarray*}
\hspace*{-15mm}\left[\1-\frac{ik\mb^2
D_0}{\ll-\lambda}-ik\mb^2\Kl\right]^{-1} & = &
\left[ [\1-ik\mb^2\Kl E_0^{-1}] E_0 \right]^{-1} \\
& = & E_0^{-1}\left[ \1-ik\mb^2\Kl E_0^{-1} \right]^{-1} .
\end{eqnarray*}
Again the matrix $\1-ik\mb^2\Kl E_0^{-1}$ has an inverse for 
$\lambda\sim\ll$ since $\Kl=\Kl^{\star}$. This gives the following
expression for the scattering matrix
\begin{eqnarray*}
\hspace*{-15mm}S(\lambda) & = & - \left[E_0^{\star}E_0^{-1}+ik\mb^2\Kl
E_0^{-1}\right] \left[\1-ik\mb^2\Kl E_0^{-1}\right]^{-1} \\
& = & - \left[E_0^{\star}E_0^{-1}+ik\mb^2\Kl E_0^{-1}\right]
\sum_{s=0}(ik\mb^2\Kl E_0^{-1})^s .
\end{eqnarray*}
Denoting $\pi_i\equiv\sqrt{\langle\phi_{0,i}|\phi_{0,i}\rangle}$ and
diagonalising we can write,
\begin{eqnarray*}
E_0^{-1} & = & \diag \left( 1-\frac{ik\mb^2\pi^2_1}{\ll-\lambda} , \ldots
 , 1-\frac{ik\mb^2\pi^2_m}{\ll-\lambda} , 1 , \ldots , 1 \right)^{-1} \\
& = & \diag \left( \frac{\ll-\lambda}{\ll-\lambda-ik\mb^2\pi^2_1} , \ldots
, \frac{\ll-\lambda}{\ll-\lambda-ik\mb^2\pi^2_m} , 1, \ldots , 1 \right) .
\end{eqnarray*}
Therefore
\begin{equation}\label{asym1}
\lim_{\lambda \rightarrow \ll}E_0^{-1} = P_0^{\perp} =\1 - P_0 .
\end{equation}
Furthermore
\begin{displaymath}
E_0^{\star}E_0^{-1} = \diag \left(
\frac{\ll-\lambda+ik\mb^2\pi^2_1}{\ll-\lambda-ik\mb^2\pi^2_1} , \ldots ,
\frac{\ll-\lambda+ik\mb^2\pi^2_m}{\ll-\lambda-ik\mb^2\pi^2_m} , 1 ,
\ldots , 1 \right) 
\end{displaymath}
which gives us the limit
\begin{equation}\label{asym2}
\lim_{\lambda \rightarrow \ll}E_0^{\star}E_0^{-1} = P_0^{\perp}
-P_0 =\1 -2P_0 .
\end{equation}
>From these limits we get 
\begin{eqnarray*}
\hspace*{-15mm}S(\ll) & = & - \left[\1 - 2P_0 + ik_0\mb^2\Kl
P_0^{\perp}\right] \sum_{s=0}(ik_0\mb^2\Kl P_0^{\perp})^s \\
& = & -\1 + 2P_0 - 2\sum_{s=1}(ik_0\mb^2P_0^{\perp}\Kl P_0^{\perp})^s \\
& = & -\1 + 2P_0 + O(\mb^2) .
\end{eqnarray*}
\vspace*{-2.5\baselineskip} \\
\hspace*{\fill} $\Box$ \\

This formula appears to imply that there may be non-zero transmission
in the case of zero connection between the rays. Actually the transmission
coefficients are not continuous with respect to
$\lambda$ uniformly in $\beta$ \cite{BMPY, MPPRY}. The physically
significant parameters of the system are obtained by averaging as functions
of $\lambda$ over the Fermi distribution so that there is no
transmission for $\beta =0$.
\begin{cor}\label{ascg}
If $\ll$ is an eigenvalue of $\cL$ such that $P_0 =\1$
then the above formula is independent of $\mb$, ie. 
\begin{displaymath}
S(\ll) = \1 
\end{displaymath}
\end{cor}
Consequently, when we have pure reflection at an eigenvalue of the
unperturbed operator, we have pure reflection regardless of the strength of
the interaction between the rays and the compact domain.

\section{Simple models}

In \cite{Mik:Pav} the authors discuss the case where $\Omega$ is the unit
disc in $R^2$ and there are four one-dimensional wires attached at the
points $\varphi= 0,\pi, \pm  {\pi/3}$. The dynamics on $\Omega$ is
given, using polar coordinates $(r,\theta)$, by the dimensionless
Schr\"{o}dinger equation
\begin{equation}
\label{dimles}
-\Delta \Psi + \left[ V_0 + \varepsilon r \cos(\theta) \right]
 \Psi =  \lambda \Psi
\end{equation}
on the domain with Neumann boundary conditions at the  boundary:
$$
\left.\frac{\partial \Psi}{\partial n }\right|_{r=1} = 0 .
$$
The dimensionless magnitude $\varepsilon$ of the governing  field is
choosen so that  the eigenfunction corresponding to the second smallest
eigenvalue has only two zeroes on the boundary of the unit circle which
divide the circumference in the ratio $2:1$. It is then easy to see that
by rotating the potential $V$ one may redirect the quantum current  from
the wire attached to the  point $\varphi = 0$ to  any other wire with
all of the other wires blocked \cite{Mik:Pav}. \\
The analysis in this case is  similar to the analysis given above except
there is now only a logarithmic singularity in the Greens function and the
Krein formula for infinite deficiency indices \cite{Kre, Nai} and
infinite-dimensional Rouchet theorem \cite{Goh:Sig} play a central
r\^{o}le. A large amount of the calculation was done using Mathematica. \\

In \cite{Har2}, using the above asymptotic formula for the scattering
matrix to choose appropriate parameters, the author discusses the case
where $\Omega$ is simply a one-dimensional ring and there is an angle of
$\pi /2$ between the rays---see figure 2. a). By applying a uniform field
to the ring, $q=0$ for the open state and $q=-3$ for the closed state, it
is easy to see that a switching effect is produced where the Fermi energy
is assumed to correspond to the smallest eigenvalue of the unperturbed
operator on the ring, ie. $\lambda_0=1$. See also \cite{ESS} where a
similar construction is considered. \\
\begin{figure}[ht]
\begin{center}
\hspace*{-10mm}\includegraphics[height=2.2in, width=6.5in]{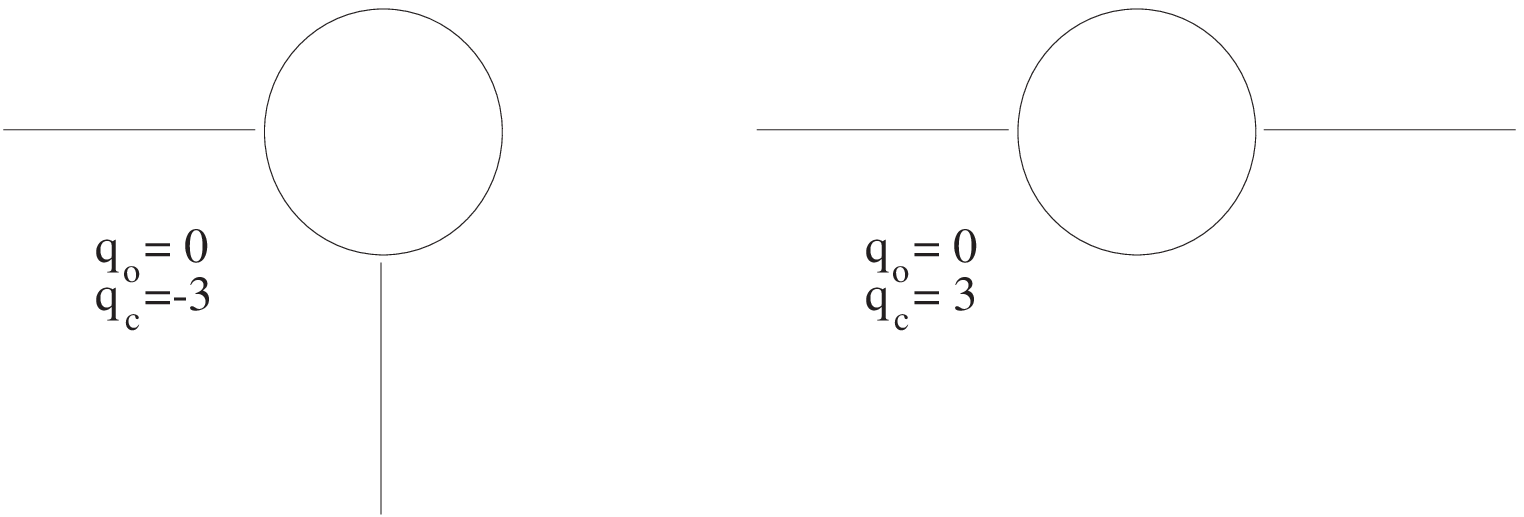} \\
 
\begin{tabular}{p{3.in}p{3.in}}
a) Interference & b) Potential barrier \\
\hspace*{15mm}switch & \hspace*{15mm}switch
\end{tabular}
figure 2. 
\end{center}
\end{figure}
Another possibility is to consider a device---see figure
2.b), the angle between the rays is now $\pi$---with similar parameters
where now we switch the current by raising a potential barrier, $q=0$ for
the open state and $q=3$ for the closed state, instead of using
interference effects. Clearly, unlike the first case, the
efficiency of such a switch will be limited by tunneling. \\  
A more detailed discussion of the properties of these two models (including plots of the averaged conductance in the closed, $\hat{\sigma}_c$, and open, $\hat{\sigma}_o$, states) is presented in \cite{Har6}. Here we just present the main observations, viz. for both switches the open state---possibly due to tunneling effects---is more difficult to achieve; and in the limit of small $\mb$, the properties of the switches improve. This is probably due to the fact that weak coupling between the ring and rays improves the open state of the switches. On the
other hand, in the limit $\mb\rightarrow 1$, the ratio $\hat{\sigma}_c
/\hat{\sigma}_o$ for the second example rapidly decreases to a bound due to
tunneling which may be calculated from the transmission coefficient
$$ 
\lim_{\tau\rightarrow 0} \left. 
\frac{\hat{\sigma}_c}{\hat{\sigma}_o} \right|_{\mb=1}\approx 4.57\times 
10^3 \, .
$$
The first switch does not have this bound and
consequently for sufficiently low temperature or small radius (see
first footnote) we conjecture that it will have better properties.


\begin{thebibliography}{10}

\bibitem{Ada}
V.~M. Adamyan.
\newblock {\em Oper. Theory: Adv. Appl.}, 59:1--10, 1992.

\bibitem{APY}
I.~Antoniou, B.~S. Pavlov, and A.~M. Yafyasov.
\newblock Quantum electronic devices based on metal-dielectric transition in
  low-dimensional quantum structures.
\newblock In D.~S. Bridges, C.~Calude, J.~Gibbons, S.~Reeves, and I.~Witten,
  editors, {\em Combinatorics, Complexity, Logic (Proceedings of DMTCS '96)},
  Singapore, 1996. Springer.

\bibitem{BMPY}
V.~Bogevolnov, A.~Mikhailova, B.~S. Pavlov, and A.~Yafyasov.
\newblock About scattering on the ring.
\newblock In A.~Dijksma, M.~A. Kaashoek, and A.~C.~M. Ran, editors, {\em Recent
  advances in operator theory (Groningen, 1998)}, pages 155--173.
  Birkh{\"{a}}user, Basel, 2001.

\bibitem{Car2}
R.~Carlson.
\newblock Inverse eigenvalue problems on directed graphs.
\newblock {\em Trans. Amer. Math. Soc.}, 351(10):4069--4088, 1999.

\bibitem{Exn}
P.~Exner.
\newblock Contact interactions on graph superlattices.
\newblock {\em J. Phys. A: Math. Gen.}, 29:87--102, 1996.

\bibitem{Exn:Seb}
P.~Exner and P.~{\u{S}}eba.
\newblock Free quantum motion on a branching graph.
\newblock {\em Rep. Math. Phys}, 28:7--26, 1989.

\bibitem{ESS}
P.~Exner, P.~{\u{S}}eba, and P.~Stovicek.
\newblock Quantum interference on graphs controlled by an external electrical
  field.
\newblock {\em J. Phys. A: Math. Gen.}, 21:4009--4019, 1988.

\bibitem{MFad}
M.~D. Faddeev.
\newblock Asymptotic behaviour of the green's function for the neumann problem
  at the boundary point.
\newblock {\em J. Sov. Math.}, 30:2336--2340, 1985.

\bibitem{Gera}
N.~I. Gerasimenko.
\newblock The inverse scattering problem on a noncompact graph.
\newblock {\em Theoret. and Math. Phys.}, 75:460--470, 1988.

\bibitem{Gera:Pav}
N.~I. Gerasimenko and B.~S. Pavlov.
\newblock Scattering problems on compact graphs.
\newblock {\em Theoret. and Math. Phys.}, 74:230--240, 1988.

\bibitem{GPP}
B.~Geyler, B.~S. Pavlov, and I.~Popov.
\newblock Possible construction of a quantum triple logic device.
\newblock New technologies for narrow-gap semiconductors, progress report (july
  1, 1998--june 30 1999), ESPRIT NTCONGS, 1999.
\newblock ESPRIT project No. 28890.

\bibitem{Goh:Sig}
I.~S. Gohberg and E.~I. Sigal.
\newblock Operator extension of the theorem about logarithmic residue and
  rouchet theorem.
\newblock {\em Mat. Sbornik.}, 84:607, 1971.

\bibitem{Har3}
M.~Harmer.
\newblock Hermitian symplectic geometry and the {Schr\"{o}dinger} operator on
  the graph.
\newblock Department of Maths Report series 444, University of Auckland, New
  Zealand, 2000.

\bibitem{Har}
M.~Harmer.
\newblock {\em The Matrix {Schr\"{o}dinger} Operator and {Schr\"{o}dinger}
  Operator on Graphs}.
\newblock PhD thesis, University of Auckland, 2000.

\bibitem{Har2}
M.~Harmer.
\newblock Scattering on the annulus.
\newblock Department of Maths Report series 445, University of Auckland, New
  Zealand, 2000.

\bibitem{Har6}
M.~Harmer, B.~S. Pavlov, and A.~Mikhailova.
\newblock Manipulating the electron current through a splitting.
\newblock {\em Proceedings of the Centre for Mathematics and its Applications},
  39:118--131, 2001.

\bibitem{Kost:Sch}
V.~Kostrykin and R.~Schrader.
\newblock Kirchhoff's rule for quantum wires.
\newblock {\em J. Phys A: Math. Gen.}, 32:595--630, 1999.

\bibitem{Kre}
M.~Krein.
\newblock {\em Comptes Rendue (Doklady) Acad. Sci. USSR (N.S.)}, 52:651--654,
  1946.

\bibitem{Kur1}
P.~B. Kurasov.
\newblock {\em J. Math. Anal. Appl.}, 201:297, 1996.

\bibitem{Mel}
Yu.~B. Melnikov.
\newblock Scattering on graphs as a quantum few-body problem.
\newblock preprint IPRT \#09-93, 1993.

\bibitem{Mel:Pav}
Yu.~B. Melnikov and B.~S. Pavlov.
\newblock Two-body scattering on a graph and application to simple
  nanoelectronic devices.
\newblock {\em J. Math. Phys}, 36:2813--2825, 1995.

\bibitem{Mik:Pav}
A.~Mikhailova and B.~S. Pavlov.
\newblock Quantum domain as a triadic relay.
\newblock Department of Maths Report series 439, University of Auckland, New
  Zealand, 2000.

\bibitem{MPPRY}
A.~Mikhailova, B.~S. Pavlov, I.~Popov, T.~Rudakova, and A.~M. Yafyasov.
\newblock Scattering on a compact domain with few semi-infinite wires attached:
  resonance case.
\newblock {\em Mathematishe Nachrichten}, 235:101--128, 2002.

\bibitem{Nai}
M.~Naimark.
\newblock {\em Bull. Acad. Sci. USSR Ser. Math.}, 4:53--104, 1940.

\bibitem{Nov3}
S.~P. Novikov.
\newblock Schr{\"{o}}dinger operators on graphs and symplectic geometry.
\newblock In E.~Bierstone, B.~Khesin, A.~Khovanskii, and J.~Marsden, editors,
  {\em The Arnol'dfest (Toronto, ON, 1997)}, volume~24 of {\em Fields Institute
  Communications}, pages 397--413, 1999.

\bibitem{Nov2}
S.~P. Novikov and I.~A. Dynnikov.
\newblock Discrete spectral symmetries of low-dimensional differential
  operators and difference operators on regular lattices and two-dimensional
  manifolds.
\newblock {\em Uspekhi Math. Nauk-Russian. Math. Surveys}, 52(5):1057--1116,
  1997.

\bibitem{Pav}
B.~S. Pavlov.
\newblock The theory of extensions and explicitly solvable models.
\newblock {\em Uspekhi Math. Nauk-Russian. Math. Surveys}, 42(6):127--168,
  1987.

\bibitem{Ree:Sim}
M.~Reed and B.~Simon.
\newblock {\em Methods of Modern Mathematical Physics}.
\newblock Academic Press, New York, 1972.

\bibitem{Soa}
P.~M. Soardi.
\newblock {\em Potential theory of infinite networks}.
\newblock Lecture Notes in Mathematics, Vol. 1590. Springer-Verlag, Berlin,
  1994.

\bibitem{YBR}
A.~M. Yafyasov, V.~B. Bogevolnov, and T.~V. Rudakova.
\newblock Quantum interferentional electronic transistor (qiet). theoretical
  analysis of electronic properties for low-dimensional systems.
\newblock (Progress Report 1995), Preprint IPRT N 99-95.

\end{thebibliography}
\end{document}